\documentclass[aps,pre,
letterpaper,
superscriptaddress,
twocolumn,
floatfix,
amsmath,amssymb,amsfonts,
]{revtex4-1}

\usepackage{graphicx}
\usepackage[breaklinks=true]{hyperref}
\usepackage{color}
\usepackage{array}
\usepackage[normalem]{ulem}

\newcolumntype{C}[1]{>{\centering\let\newline\\\arraybackslash\hspace{0pt}}m{#1}}

\begin{document}

\title{Scalable replica-exchange framework for Wang--Landau sampling}

\author{Thomas Vogel}
\email{thomasvogel@physast.uga.edu}\thanks{\\Present address: Theoretical Division, Los Alamos National
  Laboratory, Los Alamos, NM 87545, USA}
\affiliation{Center for Simulational Physics, The University of
  Georgia, Athens, GA 30602, USA}
\author{Ying Wai Li}
\affiliation{National Center for Computational Sciences, Oak Ridge
  National Laboratory, Oak Ridge, TN 37831, USA}
\author{Thomas W\"ust}
\affiliation{Scientific IT Services, ETH Z\"{u}rich, 8092 Z\"{u}rich, Switzerland}
\author{David P. Landau}
\affiliation{Center for Simulational Physics, The University of
  Georgia, Athens, GA 30602, USA}

\begin{abstract}
  We investigate a generic, parallel replica-exchange framework for
  Monte Carlo simulations based on the Wang--Landau method. To
  demonstrate its advantages and general applicability for massively
  parallel simulations of complex systems, we apply it to lattice spin
  models, the self-assembly process in amphiphilic solutions, and the
  adsorption of molecules on surfaces.  While of general, current
  interest, the latter phenomena are challenging to study
  computationally because of multiple structural transitions occurring
  over a broad temperature range.  We show how the parallel framework
  facilitates simulations of such processes and, without any loss of
  accuracy or precision, gives a significant speedup and allows for
  the study of much larger systems and much wider temperature ranges
  than possible with single-walker methods.
\end{abstract}

\maketitle

\section{Introduction}
\label{sec:intro}

In Monte Carlo simulations, one is interested in sto\-chastically
sampling the configurational space of a~model system by the creation
of a chain of consecutive micro\-states $X_i$ (via a random `walker'):
\begin{equation*}
X_0\,\stackrel{\textrm{update}}{\longrightarrow}\,
X_1\,\stackrel{\textrm{update}}{\longrightarrow}\,
\ldots\,\stackrel{\textrm{update}}{\longrightarrow}\,X_n\,.
\end{equation*}
The resultant `dynamics' is artificial and depends upon the
mechanism of proposing new microstates; $n$ is usually a~large number
$\gg 10^6$. The proposed new microstate is accepted with a certain
probability which determines the statistical ensemble. In statistical
physics, most common simulations are carried out in the canonical
ensemble by fixing the volume and particle density of a system and by
setting the acceptance probabilities equal to the fraction of
Boltzmann factors between the actual and the proposed new microstate
\cite{metropolis}.

A serious weakness of this scheme is well known: Typically, there are
barriers in the free energy landscapes of complex systems, and the
time it takes to overcome these barriers grows exponentially with
their heights. Various Monte Carlo methods have been developed to
confront the challenge of sampling such rough free energy
landscapes~\cite{muca1,muca2,sim_anneal_1,sim_anneal_2,yukito_rev} and to
carry out simulations in an ensemble where the walker is not hindered
by any barrier.  For instance, Wang--Landau
sampling~\cite{wl_prl,wl_pre} has been shown to be very effective in
overcoming energy barriers by iteratively determining the energy
density of states (DOS) of a system and seeking to perform a random
walk in energy space (`flat histogram'); i.e. eventually performing
a walk through configurational space such that all possible energies
are visited uniformly. Wang--Landau (WL) sampling was successfully
used in many scientific
problems~\cite{yamaguchi01jpa,schulz02ijmpc,yamaguchi02pre,rathore02jcp,yan02jcp,shell02pre,calvo02mp,troyer03prl,mustonen03jpa,rampf05epl,zhou06prl,strathmann08jcp,taylor09jcp,wuest09prl}.
Another powerful approach is parallel tempering or replica-exchange
Monte Carlo~\cite{partemp1,partemp2,partemp3,partemp4,elmar}. Here the
idea is to run multiple simulations in canonical ensembles at
different temperatures and propose replica or conformational exchanges
between them, which are accepted with a probability according to their
respective Boltzmann weights. This configurational mixing among
different walkers greatly alleviates the trapping problem near
conformational or energy barriers.

Whereas a parallel implementation of replica-exchange Monte Carlo is
straightforward, efficient and correct parallelization of the
Wang--Landau (WL) algorithm has posed some subtle difficulties.
Previous attempts have, for example, focused on running multiple,
independent WL samplers which simultaneously update the same density
of states~\cite{khan05jcp,zhan,yin12cpc}.  However, a recent,
massively parallel implementation of this approach~\cite{yin12cpc} has
revealed that inter-dependencies among the various WL walkers can
introduce an erroneous bias in the estimate of the DOS and thus render
this parallelization scheme highly problematic.  Another WL
parallelization, which is based on a simple splitting of the global
energy range into smaller, independently sampled, non-overlapping
energy windows, is also unsatisfactory. If the energy windows chosen
are too small, configurational space may not be sampled correctly
anymore due to ergodicity breaking, resulting again in subtle
systematic errors (this effect becomes particularly pronounced in 2D
WL simulations, where one samples a joint DOS, e.g., in energy
and magnetization space~\cite{zhou06prl}).  Moreover, the total
simulation time is bound to the WL convergence time of the slowest
walker (generally in the low energy region).

Unlike replica-exchange Monte Carlo~\cite{partemp2}, such issues have
severely limited the use of WL sampling as a means in large-scale
parallel Monte Carlo simulations. A natural route towards successful
parallelization of WL sampling would be to combine it with the
benefits of replica-exchange Monte Carlo. Variants of this general
idea have been used for specific applications by other authors on a
small scale~\cite{rathore03jcp,strathmann08jcp,nogawa11pre}, however,
no details about the implementation, parallelization, or the effect on
the performance are provided. The critical issue of combining results
from different replicas is omitted entirely.  Our
approach~\cite{letter_version} is different by introducing a fast but
generally applicable framework suitable for massive parallelization
combined with a general and precise scheme to combine results from all
walkers. We also start by running individual WL walkers in many
overlapping energy windows~\cite{wl_pre,shanho,thomaswuest_ta}
covering the whole energy range and allow for conformational exchanges
between walkers according to the actual WL weights. In addition to
splitting up the global energy range, we run multiple walkers within
each energy window. In order to avoid any possible bias, these walkers
are \emph{independent} and fulfill convergence, or flatness criteria,
individually. However, they work together by merging their weights
after each Wang--Landau iteration, reducing the statistical error
\emph{during} the simulation.

In this paper we investigate this hierarchical parallel Wang--Landau
scheme in detail with the aim to present a framework which
nevertheless remains, in a sense, as easy to implement as the original
single-walker WL method itself.  Most importantly, the scheme does not
depend on having previous knowledge about the system under
investigation. We apply the method to a standard benchmark model in
statistical mechanics, the 10-state Potts model, and to two
cutting-edge problems which have attracted great interest recently:
the self assembly of amphiphilic peptides into micelles and lipid
bilayers, and the surface adsorption process of polymers and proteins.
In Sec.~\ref{sec:method} we introduce the complete parallel
Wang--Landau framework in detail and in its most general formulation.
We discuss the key points to make the framework efficient. In
Sec.~\ref{sec:models} we describe the models which will be used in
Sec.~\ref{sec:results} to assess the applicability and accuracy of our
parallel WL scheme (\ref{sec:results_a}).  Furthermore, we present
results of performance and scaling analyses in Secs.
\ref{sec:results_b} and~\ref{sec:results_c}.  Finally, we combine our
findings and suggest further potential, methodological improvements in
Sec.~\ref{outlook}.

\section{General Parallel Framework for Wang--Landau Simulations}
\label{sec:method}

\enlargethispage{\baselineskip}

In generalized ensemble Monte Carlo simulations, one is interested in
the density of states (DOS) $g(E)$ over a large energy range $(E)$. In
our framework, $g(E)$ will be determined in parallel by employing
multiple computing cores, resulting in multiple, individual DOS
pieces. If there is no generic and precise way of putting these pieces
together in the post processing step of the simulation, the whole
framework becomes meaningless. We will hence split this section into
two parts: the production of the individual DOS pieces, and their
assembly into a global density of states. Both are equally important
aspects of our framework.

\vspace*{-5pt}
\subsection{Replica-exchange Wang--Landau sampling}

\begin{figure}
  \includegraphics[width=0.9\columnwidth]{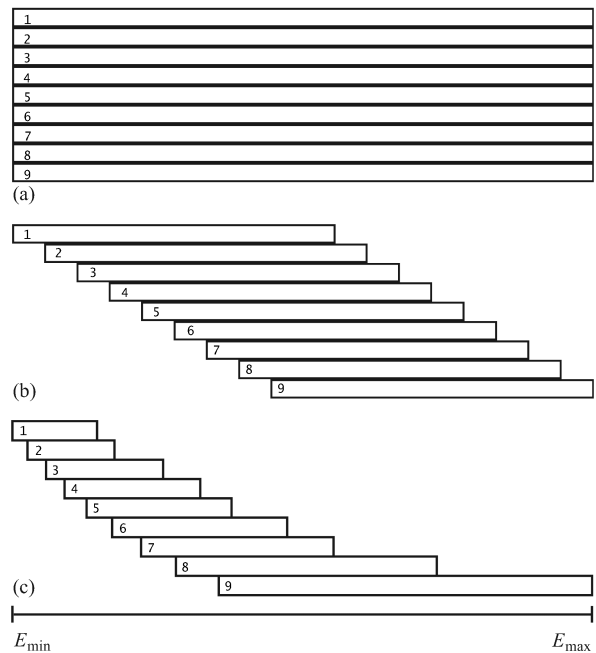}
  \caption{\label{fig:method_1}%
    Possible partitioning of the global energy range into nine
    windows: (a) the trivial case with an overlap of 100\,\%; (b)
    energy windows are slightly shifted with an overlap of 90\,\%; (c)
    an example of a run-time balanced splitting with an overlap of (at
    least) 75\,\% (note that the widths of the windows vary for
    different systems).  Several walkers might run in each window.}
\end{figure}
In the standard Wang--Landau method~\cite{wl_prl,wl_pre,shanho}, a
single walker (i.e., a chain of microstates) samples the
conformational space in an energy range between $E_\textrm{min}$ and
$E_\textrm{max}$, improving the estimate of $g(E)$ iteratively. The
microstates are sampled according to the actual weights $1/g(E)$,
which are adapted on the fly in the following way: After each Monte
Carlo trial, the estimator for $g(E(X))$ for the walker residing at
microstate $X$ is multiplied by a modification factor $f$. At the same
time the histogram of visited energies $H(E(X))$ is incremented.
Whenever there are `sufficient' entries in $H(E)$, the histogram
will be reset and the modification factor will be decreased, for
example as $\ln f\to\ln f/2.0$. Initially, $g(E) = 1.0$ and $H(E) =
0\,,\forall{}E$, and $\ln f = 1.0$. There are different approaches to
ensure that there are `sufficient' entries in the histogram for all
values of $E$, or for all energy bins, respectively.  In the original
formulation~\cite{wl_prl}, the histogram $H(E)$ is required to be
`flat', i.e.  no value for $H(E)$ may be smaller than a certain
percentage of the average histogram value.  Other
authors~\cite{zhou05pre} only insist that there are more entries in
the histogram for each $E$ than a minimal threshold number which
increases with the actual value of $f$. Additionally, a method has
been proposed where the modification factor $f$ decreases as $1/t$,
where $t$ is the simulation time~\cite{belardinelli07pre}.  (We
emphasize that our parallel framework presented here is general in the
sense that it does not depend on such details.) The estimated density
of states converges to the true one with increasing number of
iterations, and the simulation is terminated when the modification
factor reaches a minimal value $f_{\textrm{min}}$, typically set to
$\ln f_{\textrm{min}}<10^{-6} - 10^{-8}$. For all practical purposes,
the WL walker eventually performs a random walk in energy space.

For large systems, this very efficient generalized-ensemble sampling
can be enhanced by making use of multiple processors working in
parallel. This can be done, for example, by splitting up the global
energy range into smaller energy windows and estimating the density of
states for the respective windows by independent
walkers~\cite{wl_pre,shanho,thomaswuest_ta}. See
Fig.~\ref{fig:method_1} for three examples of possible energy windows.
Following the general idea of replica-exchange
methods~\cite{partemp1,partemp2,elmar}, it is natural to allow for
replica exchanges between independent walkers in WL sampling as well
if the energy windows overlap. From the detailed-balance condition for
the combined trial move, the acceptance probability $P_{\textrm{acc}}$
is derived for the exchange of conformations $X$ and $Y$ between
walkers $i$ and~$j$:
\begin{equation}
P_{\textrm{acc}}=\min\left[1,\frac{g_i(E(X))}{g_i(E(Y))}\frac{g_j(E(Y))}{g_j(E(X))}\right]\,,
\end{equation}
where $g_i(E(X))$ is the current estimator for the density of states of
walker $i$ at the energy of its present conformation (microstate $X$).

\begin{figure}
  \includegraphics[width=\columnwidth]{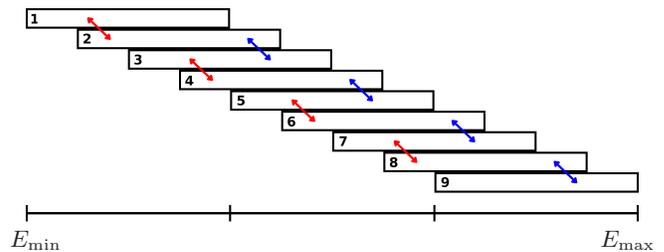}
  \caption{\label{fig:method_2}
    The communication for the replica exchange between
    energy windows alternates between the one marked by
    red (left) and the one marked by blue (right) arrows.}
\end{figure}

For replica-exchange attempts, the walker synchro\-nously communicates
with a neighbor in an adjacent energy window, see
Fig.~\ref{fig:method_2} for an illustration.  (Note, in principle one
could allow replica exchanges between walkers in any windows with
non-zero energy overlap.)  In addition to replica exchange, we will
also run multiple, independent walkers in each individual energy
window.  Each walker carries its own estimator for the density of
states $g(E)$ and its own histogram $H(E)$ and is required to fulfill
the `flatness' criterion individually. In particular, individual
histograms $H(E)$ will not add, a feature which eliminates the
potential for systematic errors as observed previously
\cite{yin12cpc}. However, walkers in the same energy window will merge
and average their individual $g(E)$ estimator before simultaneously
proceeding to the next Wang--Landau iteration step. This averaging
results in a reduction of systematic errors of $g(E)$ during the
course of the simulation and, thus, reduces the overall convergence
time. The parallel simulation terminates when every walker has reached
the final modification factor $f_{\textrm{min}}$. Walkers that have
converged to the final modification factor early will continue walking
in order to allow for replica exchange with walkers which have not
finished yet.

As in any replica-exchange scheme, one has to ensure that the exchange
of replica, and thus the overall flow of the simulation, does not get
stuck; and, hence, each replica can perform frequent round-trips over
the entire energy range. Obviously, there must be some overlap between
communicating energy windows in the first place. Furthermore, in
order to make this framework efficient, two questions arise which have
to be answered individually for different systems:

\emph{(i)} What is a reasonable overlap between neighboring energy
windows? Certainly, the overlap must not be too small or it would
lead to small acceptance rates for conformational exchanges, just like
in parallel tempering simulations when two probability distributions
have a very small overlap. On the other hand, excessive overlaps will
make the scheme inefficient, as in the extreme case one would have $n$
walkers in energy windows identical to the global energy range,
see Fig.~\ref{fig:method_1}\,a.

\emph{(ii)} What is a good number of energy windows? Generally, one
would like as many windows as possible to improve the scaling of the
simulation and exploit the capabilities of modern parallel computers.
On the other hand, when the number of energy windows increases, the
time for complete round trips also increases. Also, given a fixed
global energy range and equal-size energy windows, the number of
energy windows directly relates to their width, and too small energy
windows will eventually increase the occurrence of systematic errors,
even though the risk of `locking out' of parts of the conformational
space practically becomes negligible due to the replica-exchange
mechanism.  In any case, this problem obviously depends much more on
the actual system than the first question does.

On the algorithmic level, the scaling of the framework is important
for its general applicability and has to be investigated:

\emph{$(iii)$} How does the performance depend on the number of
computing cores used? In the optimal case, a method would achieve both
strong scaling and weak scaling, i.e., if more cores are applied, one
wants to be able to obtain results faster or to simulate larger
systems, respectively, or even both. In that context it is important
to know how the number of Wang--Landau walkers within each energy
window affects the performance of the framework and the error of the
final DOS estimate.

With the help of different `test' systems, ranging from discrete
lattice spin models to coarse-grained molecular systems in continuous
space, we will examine these questions in more detail and demonstrate
the potential of our novel approach. All models are qualitatively
different and known to show complex behavior. We will use the
following notations throughout this paper: 
 \begin{itemize}
 \item[$h$\,:] number of energy windows,
 \item[$m$\,:] number of independent walkers per energy \hbox{window},
 \item[$o$\,:] overlap with the next lowest energy window ($0\leq o\leq100\%$),
 \item[$s_h(o)$\,:] speedup of the parallel simulations with overlap $o$
   and constant $h$ (see Eq.~(\ref{eq:speed}) for \hbox{definition}),
 \item[$s_o(h)$\,:] speedup of the parallel simulations with $h$ energy
   windows at constant overlap $o$.
 \end{itemize}

\subsection{Concatenation of DOS pieces}
\label{sec:method_b}

At the end of a parallel WL simulation we are left with $h$ DOS pieces
which first need to be put together carefully before any thermodynamic
quantities, such as the internal energy or specific heat, can be
calculated. This DOS concatenation procedure is a delicate technical
challenge since even small artificial steps or kinks in the entropy
($\propto \ln[g(E)]$) may cause significant artificial peaks and oscillations
in observables like the specific heat. Therefore, simply joining DOS
pieces at some fixed positions likely results in discontinuities in
the DOS, which, in turn, cause significant artifacts in derived
quantities. To minimize such erroneous effects, we proceed as
\hbox{follows}:

(i) We calculate the first derivative $\partial\ln(g(E))/\partial E$
(i.e. the inverse microcanonical temperature) for each DOS piece in
the overlap region between two adjacent energy windows. A~large number
of overlap data points allows us to use higher order 5-point
approximations of the derivative with varying step width of the order
of up to 10 (depending on the system and the width of the energy
windows) which provide very smooth estimates for the derivatives.

(ii) We determine the point where the inverse microcanonical
temperatures of the two overlapping DOS pieces coincide the best, then
connect the DOS pieces at this point and cut the `overhanging tails'
at the respective sides of each piece, thus avoiding
non-differentiable points in the resulting entropy by construction.
This technique of connecting DOS pieces turned out to be indispensable
for obtaining accurate results as shown \hbox{below}; for more
details, see~\cite{rewlproc1,rewlproc2}.

For a rigorous error analysis, we perform $n$ independent parallel
simulations from which we get $n \times h$ individual DOS pieces. We
then calculate the mean of the $n$ DOS pieces for each energy window
before proceeding to steps (i) and (ii) above. This yields the mean
global DOS. To estimate statistical errors, we apply a bootstrap
resampling technique~\cite{newman_book}. That is, we randomly choose
$n$ DOS pieces (with repetitions) for each energy window and apply the
above two steps. This procedure is repeated multiple times (e.g. 200)
yielding multiple resampled global DOS. From these we calculate the
statistical errors of the global DOS and its derived observables. The
entire technique is unambiguous and very precise; moreover, it has an
advantage that due to the random selection of DOS pieces during the
bootstrap analysis, connection points are always at different
positions leading to statistical smoothing of potentially remaining
artifacts in derived quantities.

\section{Models and Model-Specific Algorithmic Details}
\label{sec:models}

\subsection{The Potts model for lattice spins}
\label{sec:spin_model}

The first model we use to test our parallel WL scheme is the well
studied $Q$-state Potts model for lattice spins in two
dimensions~\cite{potts1952mpcps,wu82rmp}. It is a common test bed for
novel simulation methods (see~\cite{muca2,janke93prb,wl_pre} for examples),
as the system size is scalable in a straightforward manner and exact
results exist, e.g. for the infinite size transition temperature of
the first-order phase transition between the ordered and disordered
phases for $Q>4$~\cite{baxter73jpc}. The Hamiltonian is given by
\begin{equation}
{\cal{H}}=-\sum_{\langle i,j \rangle} \delta(q_i,q_j)\,,
\end{equation}
where the spins $q_i$ can take values $q_i=1\ldots Q$ and the sum is
over all nearest neighbor pairs $\langle i,j \rangle$. For this study,
we choose the 10-state Potts model, i.e., $Q=10$, and use periodic
boundaries. The total energy range is given by $-2N\leq E\leq 0$,
where $N=L\times L$ is the total number of spins and $L$ is the
lattice size. As we use this model for demonstration purposes only, we
perform the simplest Monte Carlo update move, a random, single spin
update trial.

\subsection{The hydrophobic-polar (HP) model for protein adsorption}
\label{sec:HP_model}

The HP model~\cite{Dill1985} is a minimalist, coarse-grained lattice
model used to study generic protein folding behavior. It classifies
amino acids into only two types of monomers according to their
affinity to water: hydrophobic (H) and polar (P). There is only an
interaction between non-bonded, hydrophobic monomers occupying
nearest-neighbor sites, with a coupling strength
$\varepsilon_{\textrm{HH}}$.

Protein adsorption can be simulated with a slight modification to the
HP model~\cite{Bachmann2006}. On a three-dimensional cubic lattice, a
substrate, placed at the $z = 0$ plane, attracts H- and P-monomers in
the HP chain with a strength $\varepsilon_{\textrm{SH}}$ and
$\varepsilon_{\textrm{SP}}$, respectively. The energy function of this
model can then be written as:
\begin{equation}
  E=-\varepsilon_{\textrm{HH}}\,n_{\textrm{HH}}-\varepsilon_{\textrm{SH}}\,n_{\textrm{SH}}
  -\varepsilon_{\textrm{SP}}\,n_{\textrm{SP}}\,,
  \label{eq:hamiltonian}
\end{equation}
where $n_{\textrm{HH}}$ is the number of H--H interacting pairs and
$n_{\textrm{S[HP]}}$ is the number of H- or P-monomers adjacent to the
substrate. While periodic boundary conditions are imposed for the
$x$- and $y$-directions, a non-attractive wall is placed at $z = N +
1$ to confine the simulation box from above, where $N$ is the chain
length of the sequence.

\begin{figure}
  \includegraphics[width=.6\columnwidth]{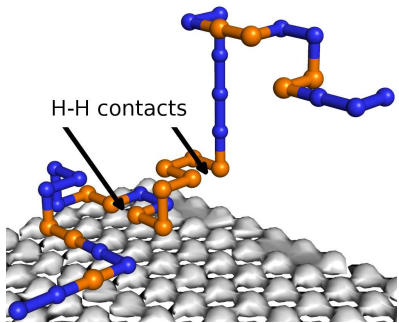}
  \caption{\label{fig:HPmodel} Illustration of a partly
    adsorbed protein composed of 36 monomers in the HP model. Besides
    the surface contacts (not shown), there are two contacts
    between non-bonded hydrophobic monomers (indicated by \hbox{arrows}).}
\end{figure}    
In the following, we consider a 36mer
(P$_3$H$_2$P$_2$H$_2$P$_5$H$_7$\allowbreak{}P$_2$H$_2$P$_4$H$_2$P$_2$HP$_2$)
\cite{Unger1993} interacting with a weakly attractive surface by
setting $\varepsilon_{\textrm{HH}} = 12$,
$\varepsilon_{\textrm{S[HP]}} = 1$.  This system has a very rugged
density of states (see below) which makes it an excellent test case
for our new parallel framework. Conformational updates are proposed by
means of pull moves and
bond-rebridging~\cite{lesh03recomb,deutsch97jcp,wuest08cpc,wuest09prl,wuest12jcp}.
For a fair comparison with previous
results~\cite{Li2011_1,Li2011_2,Li2013}, pull moves make up 20\% of
the Monte Carlo moves, while bond-rebridging moves make up 80\% of
them.  For the WL sampling, we use the originally introduced
80\%-flatness criterion~\cite{wl_prl} and the simulations end when the
modification factor reaches $\ln (f_{\textrm{min}}) = 10^{-8}$.
Fig.~\ref{fig:HPmodel} shows a schematic part of the system, for more
details \hbox{see~\cite{Li2011_1,Li2011_2,Li2013}}.

\subsection{Generic model for amphiphilic molecules\\ in solution}
\label{sec:lipid_model}

\begin{figure}
  \includegraphics[width=.5\columnwidth]{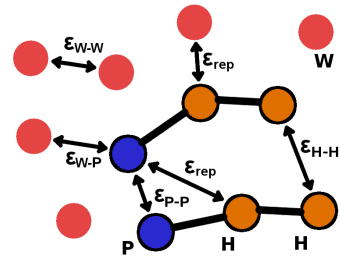}
  \caption{\label{fig:amphi.1} Illustration of particles
    and interaction parameters in the generic coarse-grained lipid
    model. P~(blue): polar head monomer, H (yellow): hydrophobic tail
    monomer, W (red): solution particle (`water molecule').}
\end{figure}

Lastly, we use a continuous, molecular model which is based on
generic models previously employed to study the self-assembly of
molecules into simple membranes, bilayers, or micelles. It includes
small amphiphilic mol\-ecules surrounded by explicit solvent
particles~\cite{getz,fujiwara}. There are three different types of
coarse-grained particles: \hbox{polar~(P)} and hydrophobic (H)
monomers and the water or solution (W) particles. The amphiphilic
molecules are composed of a polar head and two hydrophobic tail
monomers: \hbox{(P--H--H)}. The interaction between solution particles
and tail monomers as well as between head monomers and tail monomers
is purely repulsive and is modeled by a repulsive soft-core potential
\begin{equation}
  \label{eq:rep}
  U_{\textrm{repulsive}}^{\textrm{soft core}}=
  4\,\varepsilon_{\textrm{rep}}\left(\frac{\sigma_{\textrm{rep}}}{r_{ij}}\right)^{9}\,,
\end{equation}
where $\varepsilon_{\textrm{rep}}=1.0$ and
$\sigma_{\textrm{rep}}=1.05\,\sigma$ (see below),
following~\cite{getz,fujiwara}. All other non-bonded interactions are
of Lennard-Jones type
\begin{equation}
  \label{eq:lj}
  U_{\textrm{LJ}}=
  4\,\varepsilon_{X-Y}\left[\left(\frac{\sigma_{X-Y}}{r_{ij}}\right)^{12}
    -\left(\frac{\sigma_{X-Y}}{r_{ij}}\right)^{6}\right]\,.
\end{equation}
$r_{ij}$ is the Euclidean distance between two non-bonded particles
$i$ and $j$, the notation $X-Y$ stands for the interaction between
particles of type W--W, W--P, P--P, and H--H,
cf.~Fig.~\ref{fig:amphi.1}. In principle, the parameters
$\varepsilon_{X-Y}$ and $\sigma_{X-Y}$ can have different values for
different $X-Y$ combinations allowing for the introduction of
different energy scales. As we focus here on the technical aspects of
our work, we fix all $\varepsilon_{X-Y}=1.0$ and
$\sigma_{X-Y}=\sigma=2^{-1/6}\,r_0$, where $r_0=1.0$ defines the
length scale in the system. The potentials in Eqs.~(\ref{eq:rep})
and~(\ref{eq:lj}) are cut off at $r_\textrm{c}=2.5\sigma$ and shifted
such that there are no discontinuities at this point.

Bonds are modeled using the finitely extensible, non-linear elastic
(FENE) potential
\begin{equation}
U_{\textrm{bond}}^{\textrm{FENE}}=
-KR^2\ln\left[1-\left(\frac{r_{i,i+1}-r_0}{R}\right)^2\right]^{1/2}\,,
\end{equation}
where $r_{i,i+1}$ is the length of a particular bond, $R=0.3$ is half
the width of the potential (clearly, the potential diverges at
$r_{i,i+1}=r_0\pm R$) and the `spring' constant is set to $K=40$.
The equilibrium length $r_0$ at which
\hbox{$U_{\textrm{bond}}^{\textrm{FENE}}=0$} coincides with the
equilibrium length for the non-bonded potential $U_{\textrm{LJ}}$.
Periodic boundary conditions apply.

\begin{figure*}
  \includegraphics[width=.67\textwidth]{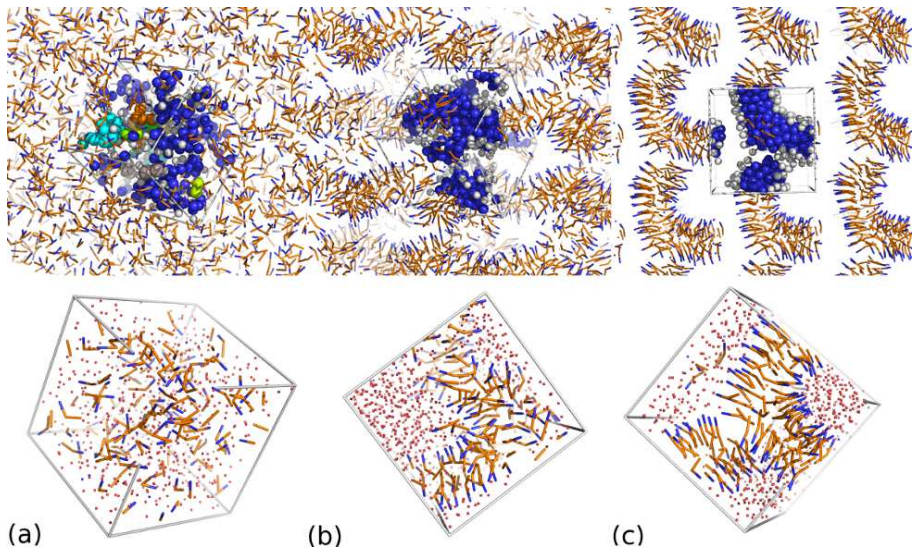}
  \caption{\label{fig:amphi.2}
    Conformations of a system containing $M=125$ amphiphilic molecules
    and a total of $N=1000$ particles. (a) Random configuration,
    $E\approx-200$ (typical initial configuration for simulation).
    (b) Amphiphilic molecules assemble and form loose clusters,
    $E\approx-4000$. (c) Low-energy, single-cluster configurations
    with compact shape, $E\approx-5100$. Upper row: Only the
    amphiphilic molecules are shown, different colors mark different
    clusters of amphiphilic molecules. Periodic copies of the
    simulated system are shown. Bottom row: Simulation boxes including
    all particles.}
\end{figure*}

Figure~\ref{fig:amphi.2} shows examples of typical configurations for
this model. The sequence of pictures shows snapshots of a system
containing $M=125$ amphiphilic molecules and $N=1000$ particles in
total at a number density of $\rho=0.8$.
One Monte Carlo (MC) sweep consists of $N$ individual MC steps. Among
these $N$ steps are, on average, $3\times M/10$ reptation moves, which
we found to be essential in order to thoroughly examine the conformation
space for large systems.  The other moves are local displacements of
individual particles. For the reptation move we first select, at
random, a solution particle in the vicinity of one end of an
amphiphilic molecule. This particle is then converted into either the
new head or the new tail of the molecule, and the opposite-end monomer
is converted into a solution particle. The bias introduced due to
different numbers of neighbors at opposite ends of an amphiphilic
molecule is accounted for in the calculation of the acceptance
probability of this move; for more details, see~\cite{rewlproc1}. For
this model, we use the originally proposed `80-percent' WL flatness
criterion, the final modification factor is
$\ln\,f_{\textrm{min}}=10^{-7}$.

\section{Results and Discussion:\\ Accuracy and Performance}
\label{sec:results}

\subsection{Number and size of energy windows;\\ degree of overlap}
\label{sec:results_a}

When applying our parallel framework to the 10-state Potts model,
(which shows a strong, temperature-driven first-order transition), we
first vary the system size ($N=1000$, $4000$, $9000$, and $16\,000$
spins) while keeping the energy window size ($\Delta E=1000$) and
overlap ($o=75\%$) fixed.  Consequently, the number of energy windows
needed to cover the whole energy range increases. We used up to nine
walkers in each energy window so that, in total, $\gtrsim10,000$ cores
were used for the biggest system.
\begin{figure*}
 \includegraphics[width=.9\textwidth]{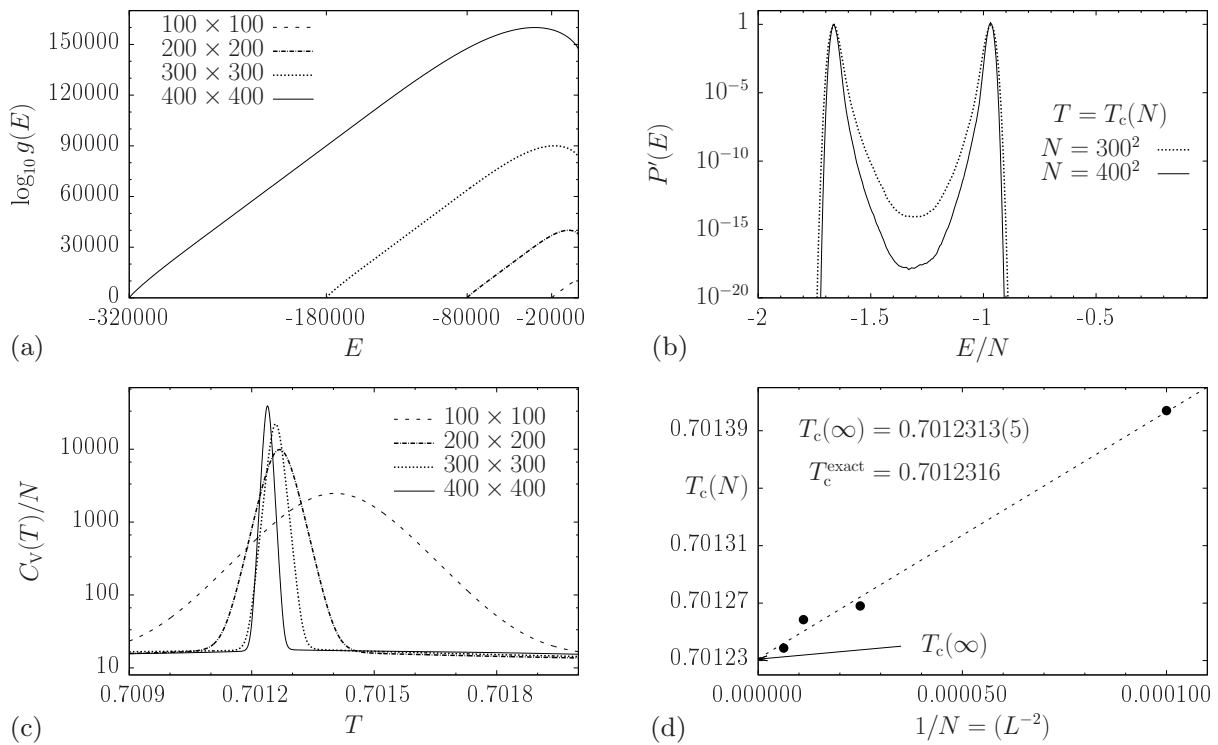}
 \caption{\label{fig:potts} a) Density of states of the 2D 10-state
   Potts model for four lattice sizes. The data are composed of 77
   ($100\times100$), 317 ($200\times200$), 717 ($300\times300$), and
   1277 ($400\times400$) pieces. b) Canonical distributions at the
   finite-size transition temperature for the larger systems. c) and
   d) The corresponding specific heat curves and the finite size
   scaling of the size-dependent transition temperature. Temperatures
   in (d) were obtained from peak positions in the specific heats in
   (c).}
\end{figure*}
In Fig.~\ref{fig:potts}a we plot the densities of states for the four
lattice sizes.  The data are composed of 77, 317, 717, and 1277 pieces
for the $N=1000$, $4000$, $9000$, and $16\,000$ spin systems,
respectively; the density of states of the largest system covers more
than $150\,000$ orders of magnitude. To further demonstrate the
simulational challenge we show in Fig.~\ref{fig:potts}b that
transition states are suppressed by a factor of $10^{-17}$--$10^{-18}$
for the largest system. To verify the results, we reproduced and
extended the corresponding analysis from the original Wang--Landau
paper~\cite{wl_pre}.  Figure~\ref{fig:potts}c shows the specific heats
for all system sizes and Fig.~\ref{fig:potts}d the finite size scaling
analysis of the transition temperature. By extrapolating the peak
position of the specific heat peaks, we estimate the transition
temperature in the thermodynamic limit to be $0.7012313(5)$ in
agreement with the exact value 0.7012316~\cite{baxter73jpc}.  While
this analysis would take years for a serial, single-walker code (the
simulation for the $100\times100$ system for a single WL walker still
takes several days), we obtained all results within a few hours
applying our parallel scheme.  The scaling analysis is presented
below.

After this first `proof of concept', we now apply our method to much
more complex molecular models. Before presenting the physical results,
we demonstrate the influence of the overlap of neighboring energy
windows on the speedup $s_h(o)$ (as compared to single-walker runs)
and acceptance rates for the replica exchange using the continuous
model for amphiphilic molecules as described in
Sec.~\ref{sec:lipid_model}. The system setup is the same as the one
used in Fig. \ref{fig:amphi.2}. We split the whole energy range of
interest into nine windows ($h=9$) and, for simplicity, employ only
one walker in each window ($m=1$). The overlaps take the following
values: $o = 50\%, 62.5\%, 75\%, 87.5\% \textrm{ or } 100\%$.
Consequently, for $h=9$, the widths of the individual energy windows
for these overlap values are $1/5$, $1/4$, $1/3$, $1/2$ (and $1$) of
the width of the global energy range, respectively, cf.
Fig.~\ref{fig:method_2} for the case ($h=9$, $o=75\%$). The speedup
$s_h(o)$ is measured by the total number of MC steps (`MC time') it
takes for the `slowest' WL walker to satisfy the termination
criterion ($t^{\mathrm{parallel}}_{\mathrm{term}}(h,o)$), compared to
that for a single-walker WL simulation
($t^{\mathrm{single}}_{\mathrm{term}}$):
\begin{equation}
\label{eq:speed}
  s_h(o)=\frac{t^{\mathrm{single}}_{\mathrm{term}}}{t^{\mathrm{parallel}}_{\mathrm{term}}(h,o)}\,.
\end{equation}
The acceptance rate $\alpha$ is the percentage of accepted
replica exchanges compared to the number of proposed exchanges.

The results of this test are shown in Table~\ref{tab:tab1}. The
statistical error of $s_h(o)$ is relatively large due to the
well-known fact that the run time of traditional WL iterations at very
small values of $f$ can vary significantly for independent runs. For
$\alpha$, minimal and maximal values are given as the acceptance rates
vary for different energy windows. Overall, the data allow us to draw
the following consistent conclusions: An overlap of $100\%$ does not
improve performance.  In fact, there is not much difference compared
to $n$ non-communicating single-walker simulations in this setup.
However, the speedup is already significant for a relatively large
overlap of $75\%$ and increases rather slowly as the overlap
decreases. On the other hand, even though acceptance rates for replica
exchange in the range of $15$--$25\%$ are satisfactory, the rates for
$o=75\%$ are much better and are comparable to the rates desired for
canonical replica-exchange simulations. Keeping in mind that more
overlap between neighboring energy windows will allow us to better
connect the individual DOS parts later, we thus decided to fix the
overlap to $o=75\%$.
\begin{table}
  \begin{ruledtabular}
    \begin{tabular}{lccccc}
      $o$ & 50\% & 62.5\% & 75\% & 87.5\% & 100\% \\
      $s_h(o)$ & $5.5\pm0.6$ & $4.4\pm0.5$ & $4.2\pm0.5$ & $2.0\pm0.3$ & $1.0\pm0.2$ \\
      $\alpha$ & 15--25\% & 25--40\% & 30--55\% & 35--75\% & 25--35\%
    \end{tabular}
  \end{ruledtabular}
  \caption{\label{tab:tab1}%
    Speedups ($s_h(o)$) and acceptance rates ($\alpha$) for replica exchange
    at different overlaps ($o$) for the amphiphilic test system.
    $h=9$, $m=1$. See text for details.
  }
\end{table}

Related to these considerations is the problem of finding a reasonable
number of energy windows. Given a fixed, global energy range as
well as a fixed overlap, the number of energy windows defines
their widths.  As mentioned before, there are conflicting goals. On
one hand, one would like to maximize the number of energy
windows, however, individual windows must not be too small or
systematic errors are induced by restricted sampling.  In order to get
reliable results, it is necessary that each replica performs walks
through the \textit{entire} energy range, which also means that
each sample walks back and forth through \textit{every} energy
window. While the round trip times might be shorter when also
allowing for replica exchange between next-nearest neighboring
windows, there would be much more technical overhead and the
implicit synchronization of the communication pattern would be much
more complex.

\begin{figure}
    \includegraphics[width=.78\columnwidth]{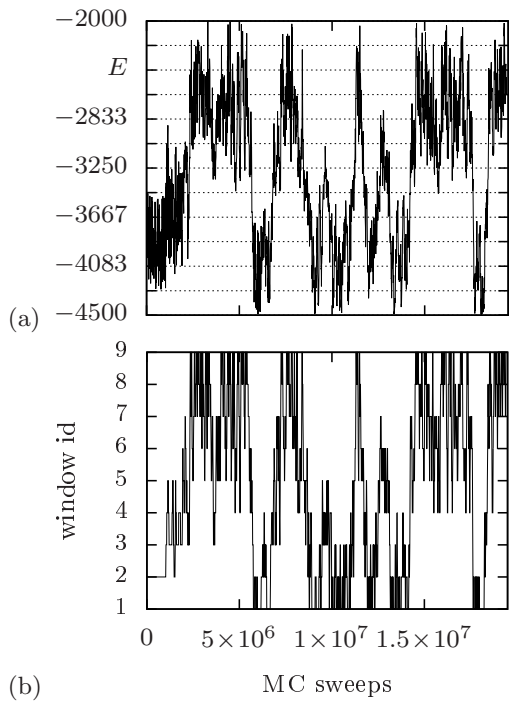}
    \caption{\label{fig:wanderlogs}%
      Path of a replica through energy space (a) and
      energy windows (b). The samples perform complete round-trips
      within less than $5\times10^6$ MC sweeps.  A conformational
      exchange between walkers is proposed every $10^4$ sweeps, with
      acceptance rates between 30 and 55\,\% (cf. Table~\ref{tab:tab1}).
      Grid lines in (a) correspond to the borders
      of the individual energy windows.}
\end{figure}
For the amphiphilic test system, we find that $h$ of the order of $10$
leads to a good performance with respect to both speedup and round
trip times for individual replica.  Eventually, for the given system
and a global energy range $E \in \left[-4500, -2000 \right]$, we find
that a splitting into nine energy windows with an overlap of 75\% is
a reasonable choice, cf.  Fig.~\ref{fig:method_2}. In
Fig.~\ref{fig:wanderlogs} we visualize the walk of a replica through
the energy space (Fig.~\ref{fig:wanderlogs}\,a) and through the
energy windows (b) during $\approx 2\times 10^7$ MC sweeps (one
MC sweep equals $N$ MC steps, where $N$ is the system size, i.e., the
total number of particles). The replica performs a smooth walk through
the energy space and completes a round trip approximately every
$5\times10^6$ MC sweeps (a replica exchange between walkers was
proposed every $10^4$ MC sweeps). We confirmed that all replicas
behaved similarly.

\begin{figure*}
  \includegraphics[width=\textwidth]{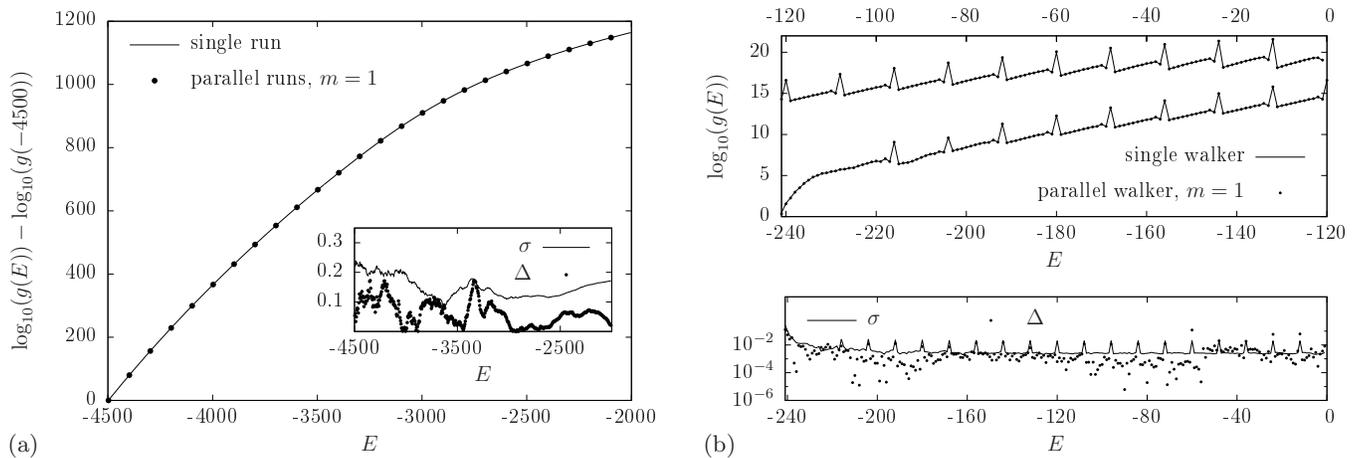}
  \caption{\label{fig:gE}%
    Logarithm of the density of states obtained
    by single-walker Wang--Landau runs (solid line) and by a
    parallel run using nine energy windows with an overlap of 75\%
    (dots). (a) Data for the system of amphiphilic molecules in
    solution. For clarity, only a small subset of data points is
    shown. (b) Data for the lattice HP 36mer on a weakly attractive
    surface. For clarity, the data are split into two curves corresponding to
    the lower and upper halves of the total energy range,
    respectively. The inset in (a) and the lower plot in (b) --
    note the logarithmic scale -- illustrate the accuracy of the
    method.  Solid lines show the standard deviation $\sigma$
    obtained from the serial, single-walker runs; dots show the
    absolute numerical difference $\Delta$ between data obtained by
    the single-walker runs and the parallel runs. All results
    obtained from the parallel runs are within the error bars of the
    reference runs.}
\end{figure*}

Figure~\ref{fig:gE} shows the logarithm of the DOS for both molecular
systems: for the continuous-energy lipid model
(cf.~Fig.~\ref{fig:gE}\,a) the global energy range is chosen to be
accessible by a single-walker WL simulation (run time of approximately
one week). For the HP lattice protein (cf.~Fig.~\ref{fig:gE}\,b) the
complete energy range ($E \in \left[-241, 0\right]$) is sampled.
Solid lines show data obtained from nine independent single-walker
runs as a reference. The statistical error bars (estimated by the
sample standard deviation~$\sigma$) are smaller than the line
thickness and are shown separately. The filled dots represent data
obtained from a single parallel run with $o=75\%$ and $h=9$ for equal
splitting energy windows.  The absolute difference between the results
from single-walker runs and the parallel run, $\Delta$, is compared to
the statistical errors from the reference (single walker) runs. For
both systems, we found that $\Delta\leq\sigma$ for practically all
energies, i.e. the results from the parallel run are clearly within
the error bars of the reference runs.  The speedup measured for the
lattice model was comparable to those shown in Table~\ref{tab:tab1}:
$s_9(75\%)=4.3\pm0.4$.

\begin{figure*}
  \includegraphics[width=\textwidth]{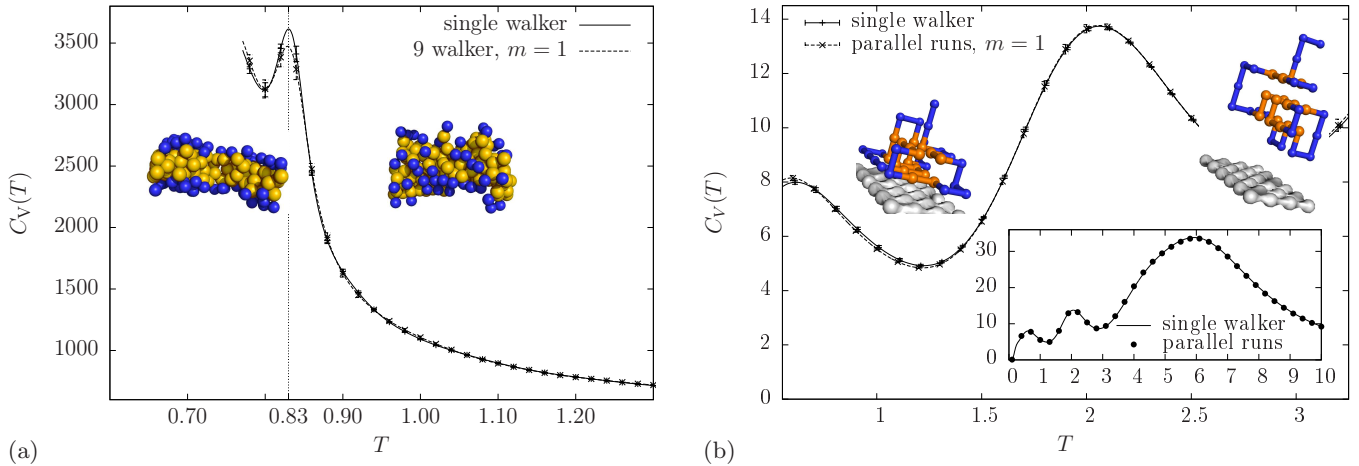}
  \caption{\label{fig:Cv}%
    Comparison of heat capacities obtained from single-walker runs
    and those from parallel runs. (a) The amphiphilic system in the region
    of the peak corresponding to the transition from cylindrical
    conformations into bilayer sections involving the alignment of
    amphiphilic molecules (for clarity, solution particles are not
    shown in pictures). (b) The 36mer in the HP model in the region of the
    adsorption peak. Pictures show representative adsorbed and
    desorbed conformations. The inset shows the same data in the
    full temperature range for reference. In both plots, the curves
    are within mutual error bars.}
\end{figure*}

\begin{figure}
  \includegraphics[width=.9\columnwidth]{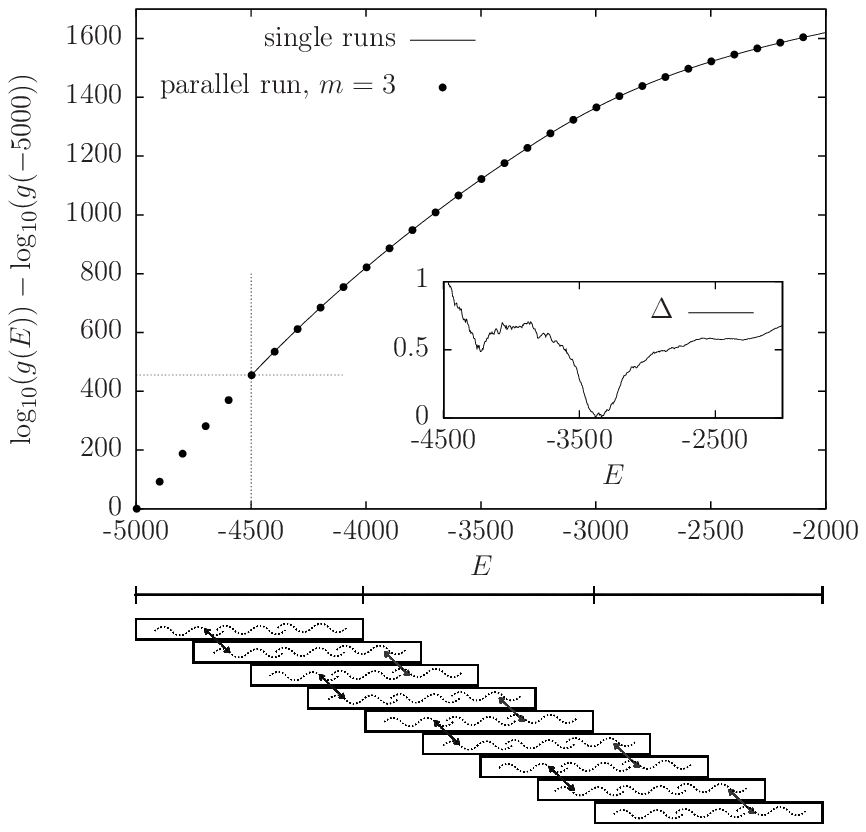}
  \caption{\label{fig:gE2amphi}%
    Logarithm of the density of states for a system
    containing $M=75$ amphiphilic molecules and a total of $N=1000$
    particles. Comparison of results from single-walker runs with
    $E_\textrm{min}=-4500$ (solid line, lower limits marked by
    dotted lines) and a parallel run with $E_\textrm{min}=-5000$,
    $m=3$ (dots). The inset shows the deviation $\Delta$ between both
    data. The diagram below the plot shows the parallel setup employed.}
\end{figure}
Sec.~\ref{sec:method_b} discussed the need for a technique to connect
DOS pieces without introducing jumps or kinks (i.e.,
non-differentiable points) into the entropy or its derivatives. In
particular, the precise results shown in Fig.~\ref{fig:gE} could not
be obtained otherwise, and kinks in the density of states would
inevitably result in artificial peaks in the heat capacity. To further
illuminate the power of our framework, in Fig.~\ref{fig:Cv} we show
parts of the heat capacities corresponding to the density of states
shown in Fig.~\ref{fig:gE}. The data clearly show that the heat
capacities obtained from the parallel and single walker (reference)
runs are within mutual error bars. For the system of amphiphilic
molecules (Fig.~\ref{fig:Cv}\,a), the small, but particularly
interesting, peak shown corresponds to the alignment of amphiphilic
molecules during the transition from cylindrical structures to liquid
bilayer sections.  (The nature of the transition was unveiled during
production runs measuring the distributions of a bond-orientation
related order parameter~\cite{seelig74biochem} and the asphericity and
prolateness resulting from the gyration
tensor~\cite{aronovitz87jpa,blavatska10jcp}). We emphasize that this
particular peak is hard to resolve in the heat capacity as it is
almost overwhelmed by stronger signals at lower temperatures (not
shown). The procedure to connect the individual DOS pieces is thus
essential to resolve and separate the signal from artifacts and
statistical noise. For further details and a study of the physics and
thermodynamic behavior of the lipid self-assembly and transitions
between different bilayer phases, see~\cite{lili_paper}. In
Fig.~\ref{fig:Cv}\,b we show the peak corresponding to the adsorption
of the 36mer HP-protein to the substrate, which lies well below the
collapse transition. See~\cite{Li2011_1,Li2011_2} for more
results and discussion.

Finally, Fig.~\ref{fig:gE2amphi} shows $\log g(E)$ for the amphiphilic
system over a larger global energy range ($E \in \left[-5000,
  -2000\right]$), which is \textit{not} accessible by single-walker
simulations anymore. The data were obtained by a parallel sampling
scheme with $h=9$, $o=75\%$ and $m=3$, hence $n=27$ walkers in total
(a sketch of this setup is given below the data). As it is not
possible to sample this energy range with a single walker, we can not
measure the speedup for this case; however, we are able to cover an
additional range of the density of states of several hundred orders of
magnitude at low temperatures (marked by dotted lines in the plot).
Owing to this extended sampling range it has been possible to uncover
the intricacies of the low temperature lipid bilayer phases such as
the liquid vs.\ gel phase -- processes which are of physiological
importance and are often studied in biochemistry and related fields;
see~\cite{lili_paper} for further discussions. In the inset of
Fig.~\ref{fig:gE2amphi}, we give the deviation $\Delta$ of the results
from the reference simulations in the energy range accessible for
such.

\subsection{Multiple walkers per energy window}
\label{sec:results_b}

\begin{figure}[b]
  \includegraphics[width=.5\columnwidth]{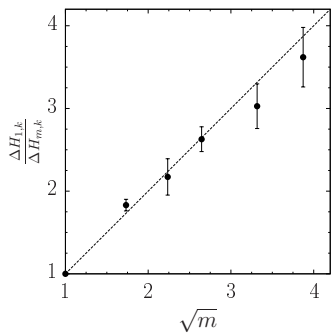}
  \caption{\label{fig:scaling} Reduction of the WL error for different
    numbers of walkers per energy window, ($m$) after convergence
    ($k = 25$) of the WL iteration. The straight dotted
    line is only a guide to the eye (not a fit to the data).}
\end{figure}
To quantify the effect of $m$ walkers per energy window, we calculate
the estimator of the error made after the $k$th WL iteration, which is
denoted by $\Delta H_{m,k}$.  We define this measure analogously to
the case for single-walker WL simulations introduced
in~\cite{lee06cpc}. In Fig.~\ref{fig:scaling} we plot $\Delta
H_{1,k=25} / \Delta H_{m,k=25}(\sqrt{m})$, i.e., the error reduction
after convergence of the WL scheme as a function of $m$.  This is
measured in the highest energy window of a run for the amphiphilic
molecules system (cf. Fig. \ref{fig:gE2amphi} for the setup). The
figure shows that the error reduces with ${\sqrt{m}}$, i.e., as for
uncorrelated WL simulations, cf. \cite{zhou05pre}. This behavior is
independent of $k$ for iterations with $\ln(f) \lesssim 10^{-2}$, i.e.
the error is reduced after each iteration during the simulation.
Furthermore, increasing $m$ can improve the convergence of the WL
procedure by reducing the risk of statistical outliers in $g(E)$,
which typically slows down subsequent iterations.

The use of multiple walkers within each energy window hence
provides the possibility to complete WL iterations faster by choosing
a weaker flatness criterion. For single-walker simulations this
usually results in faster convergence but also larger statistical
errors, but this can now be compensated for by applying multiple
walkers to reduce the error. We can, for example, apply the flatness
criterion proposed in~\cite{zhou05pre}. Then, each walker leaves its
modification factor unchanged until $\emph{all}$ walkers have
accumulated a minimum number of histogram entries for each energy
(bin), i.e., the flatness criterion is fulfilled if
\begin{equation}
H(E)\geq a/\sqrt{\ln f}\,,\quad\forall E
\label{zhoubhatt}
\end{equation}
and for all $m$ walkers inside the energy window. We thus guarantee,
that each energy has been visited by independent walkers at least
$(m\, a)/\sqrt{\ln f}$ times in total during each WL iteration, yet
every walker still fulfills the flatness criterion independently.
This ensures that systematic errors as found in~\cite{yin12cpc} cannot
occur. These $m$ walkers will then merge their DOS estimator and
proceed together to the next WL iteration. For every choice of $a$, it
should be, in principle, possible to find a value for $m$ such that
the resulting statistical error is of the same order as for a single
walker fulfilling a stricter flatness criterion. We applied this idea
to simulations of the amphiphilic model setting the parameter $a$ in
Eq.~(\ref{zhoubhatt}) as small as $1$ and measured a speedup of an
order of~10. A~\textit{quantitative} number, however, can not be given
as comprehensive simulations assuring that the statistical errors are
of equal size for both approaches would require unreasonable
computational costs just for this purpose. (This approach shares basic
ideas with a recently proposed technique of merging histograms in
multicanonical simulations~\cite{zierenberg13cpc}.)

A setup with multiple walkers per energy window is also suitable for
fault tolerant implementations~\cite{knoxville-coop} of the simulation
code. Besides the fact that hardware or network failures otherwise
usually result in a complete abortion of the program, a loss of some
walkers due to such failures can easily be tolerated if there are
multiple walkers in each window. In most cases, it would not
`disconnect' the `communication' between the outer energy windows but
only affect the statistical error in that region where walkers fail.

\subsection{Strong and weak scaling behavior}
\label{sec:results_c}

\begin{figure}
  \includegraphics[width=\columnwidth]{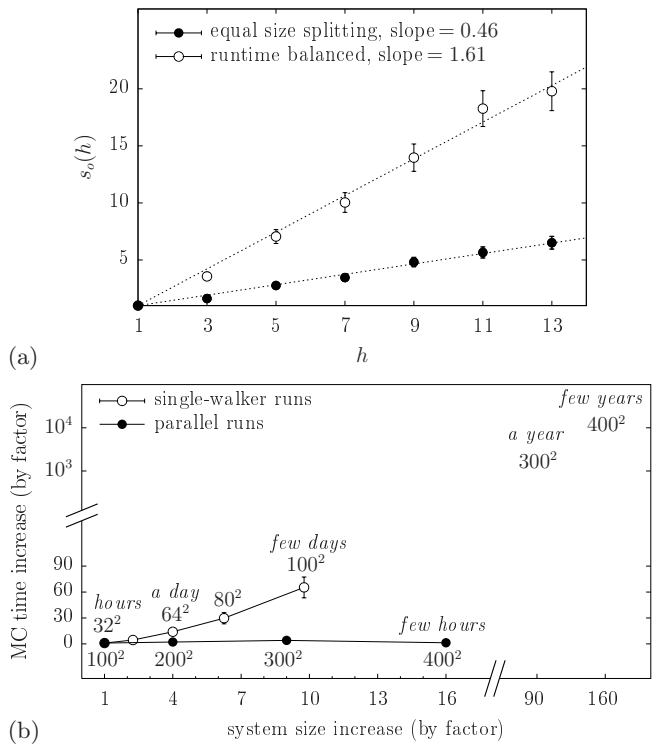}
  \caption{\label{fig:scaling_a} Scaling results: (a) Speedups
    $s_o(h)$ for different numbers of energy windows $h$ of equal size
    and overlap $o = 75\%$ (filled circles, cf.
    Fig.~\ref{fig:method_2}) and using run-time balanced energy
    splitting (open circles, cf. Fig. \ref{fig:method_1}\,c), measured
    using the continuous lipid model. $h = 1$ corresponds to the
    single walker (one CPU) reference runs. The speedup is determined
    by the MC sweeps needed to complete the first WL iteration.
    Straight dotted lines are fits to the data. (b) Simulation-time
    increase for increasing system size for the 10-state Potts model.
    Results for single-walker runs (open symbols); parallel runs
    (filled symbols).  }
\end{figure}

We now consider the scaling of our method with increasing number of
computing cores including a detailed analysis of both the speedup
(strong scaling) and the potential for simulating larger system size
(weak scaling) resulting from the increase of the number of individual
energy windows. In Fig.~\ref{fig:scaling_a}\,a we show the speedup
$s_{75\%}(h)$ for different numbers of energy windows $h\lesssim15$ in
the first WL iteration while keeping the number of processes in each
energy window fixed.  We employ $m=1$ walker per energy window and use
the continuous, amphiphilic system for this test. We find that the
speedup scales linearly with the number of energy windows in this
region, whereas the slope depends on the method of splitting of the
energy range. With a run-time balanced speedup~\cite{letter_version}
we can achieve a slope~$>1$, i.e., a speedup which is greater than the
increase in the number of cores used.

Note again that for our purpose we are using a particular definition
of speedup, which might differ from others commonly used, and that
this surprising result can be attributed to a combination of two
effects.  First, due to the way the WL algorithm behaves when building
up the entropy estimator (see, e.g.,~\cite{zhou05pre}), the total
effort (e.g. MC sweeps) needed for multiple walkers to fill up their
individual histograms can, depending on the energy splitting and the
shape of the entropy curve, be less than the work needed for a single
walker to fill a histogram over the whole energy range. Second, there
could be an algorithmic speedup through the introduction of the
replica exchange move, an effect which is independent of the speedup
from parallelization using multiple physical CPU cores. Indeed, timing
experiments on a moderate-sized ($48\times48$) Potts model using 15
energy windows revealed such an algorithmic speedup of up to roughly a
factor of 2, depending on the frequency of the replica exchange move,
compared to a scheme without replica exchanges (data not shown).
Hence, not only does the replica exchange move between multiple energy
windows reduce systematic errors in the WL simulations, it can
potentially increase the algorithmic efficiency.

However, we stress that the actual performance and the relative
importance of different mechanisms would vary from model to model;
they are also related to the settings of the simulations. The speedup
results presented above are merely examples and should not be taken as
universal values for our framework.

In Fig.~\ref{fig:scaling_a}\,b we show the growth of simulation time
with system size. We use the 2D Potts model here as the scaling of the
system is straightforward and the corresponding increase of the global
energy range is known. We compare single-walker runs (open symbols)
with parallel runs (filled symbols) where we increase the number of
energy windows correspondingly with the increase in system size. Hence
the size of individual energy windows and the overlap are fixed and we
add windows as needed to cover the entire global energy range. By
doing this we can keep the simulation time practically constant
compared to the increase in simulation time for single-walker runs.

In summary, we find both strong and weak scaling: (i) for a given
global energy range, a small number of additional energy windows can
increase the speedup significantly, and (ii) system sizes can be
extended without increasing the run time significantly by introducing
additional individual energy windows.

\section{Summary and Perspectives}
\label{outlook}

We thoroughly investigated the properties of our recently introduced
parallel framework for replica-exchange Wang--Landau sampling
(REWL)~\cite{letter_version}. The basic idea is to restrict individual
WL walkers to small, but overlapping, energy windows and enable them
to communicate with neighbors such that replica of the system can
travel through the whole energy space. In contrast to traditional
replica exchange Monte Carlo for which an unfortunate choice of
temperatures leads to little overlap of probability distributions and
effectively eliminates exchange of replicas, our scheme insures the
possibility of replica exchange by fixing overlap of energy windows at
the outset. We demonstrated the strength of the simulation framework
via a sophisticated data analysis procedure to connect the resulting
pieces of the density of states. We were able to reproduce very
accurate and precise results of single-walker simulations, only much
faster, and to facilitate simulations of systems which were not at all
accessible before. By applying the proposed framework to qualitatively
very different, challenging models, we showed that the method is
generally applicable and robust.  While it is possible to reduce the
systematic errors by employing multiple walkers in an energy window,
we also demonstrated that the proposed method shows both weak and
strong scaling when increasing the number of computing cores.  This
parallel framework is much more efficient than single-walker WL
sampling even on small hardware architectures like multi-core CPUs
but, more importantly, it can be readily implemented on larger
systems, potentially with $>10^5$ processors, by making use of all
lines of parallelization presented here.\looseness1

The description of our method was held intentionally as simple and
general as possible. Hence, despite its proven advantages in its
current form, the procedure leaves much room for further optimization.
Potential improvements are conceivable by fine-tuning energy windows
sizes and overlaps, frequency of replica-exchange moves, special
treatment at window boundaries, etc.  It is obvious, for example, that
walkers in different equal-size energy windows will not proceed
through the WL scheme at equal pace, i.e., the time it takes to
fulfill the flatness criteria will greatly differ, particularly in the
first iterations when the estimator for the density of states still
differs significantly from the true DOS~\cite{letter_version}.  At
that stage, the shape of the histogram $H(E)$ is not really flat.
Hence, an optimal energy window size distribution would be based on
the area under the respective local histograms such that the flatness
criteria in each window will be fulfilled after approximately the same
number of MC steps. On the other hand, in later stages of the
iteration, when the walkers perform almost random walks through energy
space, equal size energy windows might be indeed favorable.
Eventually, one would like to end up with self-tuning, variable-size
energy windows such that all walkers proceed synchronously through the
parallel WL scheme at all times.

\begin{acknowledgments}
  This work has been supported by the National Science Foundation
  under Grants DMR-0810223 and OCI-0904685. Y.\,W. Li was partly
  sponsored by the Office of Advanced Scientific Computing Research;
  U.\,S.  Department of Energy. Part of the work was performed at the
  Oak Ridge Leadership Computing Facility at ORNL, which is managed by
  UT-Battelle, LLC under Contract No. De-AC05-00OR22725. Supercomputer
  time was provided by TACC under XSEDE grant PHY130009.
  LA-UR-13-29579 assigned.
\end{acknowledgments}


\begin{thebibliography}{99}

\bibitem{metropolis} N. Metropolis, A. W. Rosenbluth, M. N.
  Rosenbluth, A. H. Teller, and E. Teller, J. Chem. Phys. \textbf{21},
  1087 (1953).

\bibitem{muca1} B. A. Berg and T. Neuhaus, Phys. Lett. B
  \textbf{267}, 249 (1991).

\bibitem{muca2} B. A. Berg and T. Neuhaus, Phys. Rev. Lett.
  \textbf{68}, 9 (1992).

\bibitem{sim_anneal_1} A. P. Lyubartsev, A. A. Martsinovski, S. V.
  Shevkunov, and P. N. Vorontsov-Velyaminov, J. Chem. Phys.
  \textbf{96}, 1776 (1992).

\bibitem{sim_anneal_2} E. Marinari and G. Parisi, Europhys. Lett.
  \textbf{19}, 451 (1992).

\bibitem{yukito_rev} Y. Iba, Int. J. Mod. Phys. C \textbf{12}(5),
  623 (2001).

\bibitem{wl_prl} F. Wang and D. P. Landau, Phys. Rev. Lett.
  \textbf{86}, 2050 (2001).

\bibitem{wl_pre} F. Wang and D. P. Landau, Phys. Rev. E \textbf{64},
  056101 (2001).

\bibitem{yamaguchi01jpa} C. Yamaguchi and Y. Okabe, J. Phys. A: Math.
  Gen. \textbf{34}, 8781 (2001).

\bibitem{schulz02ijmpc} B. J. Schulz, K. Binder, and M. M\"{u}ller,
  Intl. J. Mod. Phys. C \textbf{13}, 477 (2002).

\bibitem{yamaguchi02pre} C. Yamaguchi and N. Kawashima, Phys. Rev. E
  \textbf{65}, 056710 (2002).

\bibitem{rathore02jcp} N. Rathore and J. J. de Pablo, J. Chem. Phys.
  \textbf{116}, 7225 (2002).

\bibitem{yan02jcp} Q. Yan, R. Faller, and J. J. de Pablo, J. Chem.
  Phys. \textbf{116}, 8745 (2002).

\bibitem{shell02pre} M. S. Shell, P. G. Debenedetti, and A. Z.
  Panagiotopoulos, Phys. Rev. E \textbf{66}, 056703 (2002).

\bibitem{calvo02mp} F. Calvo, Mol. Phys. \textbf{100}, 3421 (2002).

\bibitem{troyer03prl} M. Troyer, S. Wessel, and F. Alet, Phys. Rev.
  Lett. \textbf{90}, 120201 (2003).

\bibitem{mustonen03jpa} V. Mustonen and R. Rajesh, J. Phys. A: Math.
  Gen. \textbf{36}, 6651 (2003).

\bibitem{rampf05epl} F. Rampf, W. Paul, and K. Binder, Europhys. Lett.
  \textbf{70}, 628 (2005).

\bibitem{zhou06prl} C. Zhou, T. C. Schulthess, S. Torbr\"{u}gge, and
  D. P. Landau, Phys. Rev. Lett. \textbf{96}, 120201 (2006).

\bibitem{strathmann08jcp} J. Luettmer-Strathmann, F. Rampf, W. Paul,
  and K.~Binder, J. Chem. Phys. \textbf{128}, 064903 (2008).

\bibitem{taylor09jcp} M. P. Taylor, W. Paul, and K. Binder, J. Chem.
  Phys. \textbf{131}, 114907 (2009).

\bibitem{wuest09prl} T. W\"{u}st and D. P. Landau, Phys. Rev. Lett.
  \textbf{102}, 178101 (2009).

\bibitem{partemp1} C. J. Geyer, in Computing Science and Statistics:
  Proceedings of the 23rd Symposium on the Interface, ed. by E. M.
  Keramidas (Interface Foundation, Fairfax Station, VA, 1991), p. 156.

\bibitem{partemp2} K. Hukushima and K. Nemoto, J. Phys. Soc. Jpn.
  \textbf{65}, 1604 (1996).

\bibitem{partemp3} D. J. Earl and M. W. Deem, Phys. Chem. Chem. Phys.
  \textbf{7}, 3910 (2005).

\bibitem{partemp4} H. G. Katzgraber, S. Trebst, D. A. Huse, and M.
  Troyer, J. Stat. Mech. (JSTAT) \textbf{2006}, P03018 (2006).

\bibitem{elmar} E. Bittner, A. Nu{\ss}baumer, and W. Janke, Phys. Rev.
  Lett. \textbf{101}, 130603 (2008).

\bibitem{rathore03jcp} N. Rathore, T. A. Knotts, and J. J. de Pablo,
  J. Chem. Phys. \textbf{118}, 4285 (2003).

\bibitem{khan05jcp} M. O. Khan, G. Kennedy, and D. Y. C. Chan, J.
  Comput. Chem. \textbf{26}, 72 (2005).

\bibitem{zhan} L. Zhan, Comput. Phys. Commun. \textbf{179}, 339 (2008).

\bibitem{yin12cpc} J. Yin and D. P. Landau, Comput. Phys. Commun.
  \textbf{183}, 1568 (2012).


\bibitem{letter_version} T. Vogel, Y. W. Li, T. W\"{u}st and D. P.
  Landau,
  \href{http://dx.doi.org/10.1103/PhysRevLett.110.210603}{Phys. Rev.
    Lett. \textbf{110}, 210603 (2013)}.

\bibitem{shanho} D. P. Landau, S.-H. Tsai, and M. Exler, Am. J. Phys.
  \textbf{72}, 1294 (2004).

\bibitem{thomaswuest_ta} T. W\"{u}st and D. P. Landau (to appear)

\bibitem{nogawa11pre} T. Nogawa, N. Ito, and H. Watanabe, Phys. Rev. E
  \textbf{84}, 061107 (2011).

\bibitem{zhou05pre} C. Zhou and R. N. Bhatt, Phys. Rev. E, \textbf{72},
  025701(R) (2005).

\bibitem{belardinelli07pre} R. E. Belardinelli and V. D. Pereyra, Phys. Rev.
  E \textbf{75}, 046701 (2007).

\bibitem{rewlproc1} T. Vogel, Y. W. Li, T. W\"ust, and D. P. Landau,
  \href{http://dx.doi.org/10.1088/1742-6596/487/1/012001}{J. Phys.: Conf. Ser. \textbf{487}, 012001 (2014)}.

\bibitem{rewlproc2} Y. W. Li, T. Vogel, T. W\"ust, and D. P. Landau,
  \href{http://dx.doi.org/10.1088/1742-6596/510/1/012012}{J. Phys.: Conf. Ser. \textbf{510}, 012012 (2014)}.

\bibitem{newman_book} M. E. J. Newman und G. T. Barkema, \textit{Monte
    Carlo methods in statistical physics}, Oxford University Press
  (Oxford, New York, 1999), and references therein.

\bibitem{potts1952mpcps} R. B. Potts, Math. Proc. Camb. Phil. Soc.
  \textbf{48}, 106 (1952).

\bibitem{wu82rmp} F. Y. Wu, Rev. Mod. Phys. \textbf{54}, 235 (1982).

\bibitem{janke93prb} W. Janke, Phys. Rev. B \textbf{47}, 14757 (1993).

\bibitem{baxter73jpc} R. J. Baxter, J. Phys. C: Solid State Phys.
  \textbf{6}, L445 (1973).

\bibitem{Dill1985} K. A. Dill, Biochemistry \textbf{24}, 1501 (1985).

\bibitem{Bachmann2006} M. Bachmann and W. Janke, Phys. Rev. E
  \textbf{73}, 020901(R) (2006).

\bibitem{Unger1993} R. Unger and J. Moult, J. Mol. Biol. \textbf{231},
  75 (1993).

\bibitem{lesh03recomb} N. Lesh, M. Mitzenmacher, and S. Whitesides,
  RECOMB '03 Proc. 7th Ann. Int. Conf. Res. Comp. Mol. Biol., pp.
  188--195.

\bibitem{deutsch97jcp} J. M. Deutsch, J. Chem. Phys. \textbf{106},
  8849 (1997).

\bibitem{wuest08cpc} T. W\"ust and D. P. Landau, Comput. Phys. Commun.
  \textbf{179}, 124 (2008).

\bibitem{wuest12jcp} T. W\"ust and D. P. Landau, J. Chem. Phys.
  \textbf{137}, 064903 (2012).

\bibitem{Li2011_1} Y. W. Li, T. W\"ust and D. P. Landau, Comput. Phys.
  Commun. \textbf{182}, 1896 (2011).

\bibitem{Li2011_2} T. W\"ust, Y. W. Li, and D. P. Landau, J. Stat. Phys.
  \textbf{144}, 638 (2011).

\bibitem{Li2013} Y. W. Li, T. W\"ust and D. P. Landau, Phys. Rev. E
  \textbf{87}, 012706 (2013).

\bibitem{getz} R. Goetz and R. Lipowsky, J. Chem. Phys. \textbf{108},
  7397 (1998).

\bibitem{fujiwara} S. Fujiwara, T. Itoh, M. Hashimoto, and R.
  Horiuchi, J.~Chem. Phys. \textbf{130}, 144901 (2009).


\bibitem{seelig74biochem} A. Seelig and J. Seelig, Biochem.
  \textbf{13}, 4839 (1974).

\bibitem{aronovitz87jpa} J. A. Aronovitz and M. J. Stephen, J. Phys.
  A: Math. Gen. \textbf{20}, 2539 (1987).

\bibitem{blavatska10jcp} V. Blavatska and W. Janke, J. Chem. Phys.
  \textbf{133}, 184903 (2010).

\bibitem{lili_paper} T. Vogel, L. Gai, C. McCabe, P. T. Cummings, D.
  P. Landau, J. Chem. Phys. \textbf{139}, 054505 (2013).

\bibitem{lee06cpc} H. K. Lee, Y. Okabe, and D. P. Landau, Comput. Phys.
  Commun. \textbf{175}, 36 (2006).

\bibitem{zierenberg13cpc} J. Zierenberg, M. Marenz, and W. Janke,
  Comput. Phys. Commun. \textbf{184}, 1155 (2013).

\bibitem{knoxville-coop} W. Bland, A. Bouteiller, T. Herault, J.
  Hursey, G.~Bosilca, and J. J. Dongarra, in J. L. Tr{\"a}ff, S.
  Benkner, and J. J. Dongarra (eds.) 19th EuroMPI, Proceedings. LNCS
  vol. 7490 (Springer, Berlin, 2012), pp. 193--203.

\end{thebibliography}

\end{document}